\documentclass[12pt]{iopart}
\usepackage{graphicx}
\begin{document}
\title[Charmed meson and charmonium production in PbPb collisions at the LHC]{Charmed meson and charmonium production in PbPb collisions at the LHC}

\author{I P Lokhtin$^1$, A V Belyaev$^1$, G Kh Eyyubova$^1$,
G Ponimatkin$^2$ and E Yu Pronina$^1$}

\address{$^1$ Lomonosov Moscow State University, Skobeltsyn Institute of Nuclear Physics,
	Moscow, Russia}
\address{$^2$ Czech Technical University in Prague, Faculty of Nuclear Sciences and
Physical Engineering, Prague, Czech Republic}

\ead{Igor.Lokhtin@cern.ch}

\begin{abstract}
The phenomenological analysis of various characteristics of $J/\psi$ and $D$ meson production in PbPb 
collisions at the center-of-mass energy 2.76 TeV per nucleon pair is presented.  
The data on momentum spectra and elliptic flow are reproduced by two-component model HYDJET++ including thermal and 
non-thermal charm production mechanisms. The significant part of 
$D$-mesons is found to be in a kinetic equilibrium with the created medium, while $J/\psi$-mesons 
are characterized by earlier (as compared to light hadrons) freeze-out.
\end{abstract}

\submitto{\jpg}
\maketitle

\section{Introduction}
Heavy quarks are predominantly produced in hard scatterings on a short time-scale and
traverse the surrounding medium interacting with its constituents. Thus the production 
of hadrons containing heavy quark(s) is a particularly useful tool to probe transport  
properties of hot matter formed in ultrarelativistic heavy ion collisions.
The modern pattern of multi-particle production in central heavy ion collisions at 
RHIC and LHC energies supposes the formation of hot strongly-interacting matter with 
hydrodynamical properties (``quark-gluon fluid''), which absorbs energetic 
quarks and gluons due to their multiple scattering and medium-induced energy loss 
(see, e.g.,~\cite{brahms,phobos,star,phenix,Armesto:2015ioy}). Within such paradigm, a number of 
questions related to heavy 
flavor production is definitely of interest. Are heavy quarks thermalized in quark-gluon plasma? 
Are charmed hadrons and charmonia in a kinetic equilibrium with the created medium? What is the interplay 
between thermal and non-thermal mechanisms of hidden and open charm production? 

Interesting measurements at the LHC involving momentum and centrality dependencies 
of charmed meson and charmonium production and its azimuthal anisotropy in PbPb collisions at center-of-mass 
energy 2.76 TeV per nucleon pair have been done by 
ALICE~\cite{Abelev:2012rv,ALICE:2012ab,Abelev:2013lca,ALICE:2013xna,Abelev:2013ila,Abelev:2014ipa,Adam:2015rba,Adam:2015nna,Adam:2015isa,Adam:2015sza,Adam:2015jda,Adam:2015gba} 
ATLAS~\cite{Aad:2010aa} and CMS~\cite{Chatrchyan:2012np,Khachatryan:2014bva} Collaborations. At that a number 
of theoretical calculations and Monte-Carlo simulations in different approaches were attempted to reproduce these 
data~\cite{Andronic:2011yq,Uphoff:2012gb,Zhao:2012gc,He:2012xz,Nahrgang:2013xaa,Zhou:2014kka,Cao:2014dja,Djordjevic:2014tka,Beraudo:2014boa,Saraswat:2015ena,Bratkovskaya:2015foa}. 
Note that the simultaneous description of momentum spectra (nuclear modification factors) and elliptic 
flow coefficients of charmed mesons is currently a challenging problem for most theoretical models.
In this paper, the LHC PbPb data on momentum spectra and elliptic flow of charmed mesons 
($D^{\pm}$, $D^{*\pm}$, $D^0$)
and $J/\psi$ mesons are analyzed 
and interpreted within two-component HYDJET++ model~\cite{Lokhtin:2008xi}. Among other 
heavy ion event generators, HYDJET++ focuses on the simulation of the jet quenching effect 
taking into account medium-induced radiative and collisional partonic energy loss (hard 
"non-thermal" component), and reproducing the main features of nuclear
collective dynamics by the parametrization of relativistic hydrodynamics
with preset freeze-out conditions (soft "thermal" component). It has been shown in the previous 
papers~\cite{Lokhtin:2012re,Bravina:2013xla,Eyyubova:2014dha,Bravina:2015sda} that the   
model is able to reproduce the LHC data on various physical observables measured in PbPb collisions at $\sqrt{s_{\rm NN}}=2.76$ TeV, 
such as centrality and pseudorapidity dependence of inclusive charged particle multiplicity, transverse momentum 
spectra of inclusive and identified ($\pi$, $K$, $p$) hadrons, $\pi^{\pm}\pi^{\pm}$ femtoscopic correlation radii, momentum 
and centrality dependencies of elliptic and higher-order harmonic coefficients, dihadron angular correlations and 
event-by-event fluctuations of anisotropic flow. The next step is to apply this model for phenomenological 
analysis of LHC data on open ($D$ mesons) and hidden ($J/\psi$ mesons) charm production. 

\section{Simulation of charm production in HYDJET++ model}

HYDJET++ is a model of relativistic heavy ion collisions which includes two independent 
components: the soft hydro-type state (``thermal'' component) and the hard state resulting from 
the medium-modified multi-parton fragmentation (``non-thermal'' component). When the Monte-Carlo 
generation of both components in each heavy ion collision is completed, the overall final particle 
spectrum is formed by natural way as the junction of these two independent event outputs. The details 
of the model and corresponding simulation procedure can be found in the HYDJET++ 
manual~\cite{Lokhtin:2008xi}. Main features of the model are listed below as follows.

\subsection {Soft component} 

The soft component represents the hadronic state generated on the chemical and thermal freeze-out 
hypersurfaces obtained from the pa\-ra\-met\-ri\-za\-ti\-on of relativistic hydrodynamics with preset 
freeze-out conditions (the adapted event generator FAST MC~\cite{Amelin:2006qe,Amelin:2007ic}). 
It is supposed that a hydrodynamic expansion of the fireball ends by a sudden system breakup 
(``freeze-out'') at given temperature $T$. Thermal production of charmed hadrons is 
treated within the statistical hadronization approach~\cite{Andronic:2003zv,Andronic:2006ky}. The momentum 
spectrum of produced hadrons retains the thermal character of the (partially) equilibrated 
Lorentz invariant distribution function in the fluid element rest frame: 
\begin{equation}
f_{\rm c}(p^{*0};T,\gamma_{\rm c}) = \frac{\gamma_{\rm c}^{\rm n_c} g_i}
{\exp{(p^{*0}/T)} \pm 1}~, 
\end{equation}
where $p^{*0}$ is the hadron energy in the fluid element rest frame, 
$g_i=2 J_i+1$ is the spin degeneracy factor, $\gamma_c \ge 1$  is the charm enhancement factor 
(or charm fugacity), and $n_{\rm c}$ is the number of charm quarks and antiquarks in a hadron $C$
($C = D, J/\psi, \Lambda_c$).  
The signs $\pm$ in the denominator account for the quantum statistics of a fermion or a boson, 
respectively. The fugacity $\gamma_c$ takes into account the enhanced yield of charmed hadrons and charmonia as 
compared with its thermal number. Note that the recombination of $c \bar{c}$ pairs to 
$J/\psi$-mesons during the hadronization stage is effectively taken into account within such an 
approach. The fugacity can be treated as a free parameter of the model, or calculated 
through the number of charm quark pairs obtained from perturbative QCD and multiplied by the 
number of nucleon-nucleon sub-collisions. 

The mean charmed hadron and charmonium multiplicities $\overline{N_{\rm c}}$ are determined 
through the corresponding thermal numbers using the effective volume approximation:
\begin{equation}
\label{Nc}
\overline{N_{\rm c}}=\rho_{\rm c}^{\rm eq}(T) V_{\rm eff}~,~~~\rho_{\rm c}^{\rm eq}(T)=\int d^3 p^{*} 
f_{\rm c}(p^{*0};T,\gamma_{\rm c})~,
\end{equation}
where $\rho_{\rm c}^{\rm eq}(T)$ is the thermal (equilibrium) density of hadrons of type $C$ at the 
temperature $T$, and $V_{\rm eff}$ is the total effective volume of hadron emission from the hypersurface 
of proper time $\tau$=const. The latter is calculated at given impact parameter $b$ of a heavy ion collision as
\begin{eqnarray}
V_{\rm eff} & = & \tau \int\limits_0^{2\pi} d\phi \int \limits_0^{R(b,\phi)}  
 \sqrt{1+\delta(b) \tanh^2 Y_{\rm T} (r,b) \cos 2 \phi} \nonumber \\ \label{Veff}
& & \cosh Y_{\rm T} (r,b) r dr \int\limits_{\eta_{\min}}^{\eta_{\max}} Y_{\rm L}(\eta) d\eta,
\end{eqnarray}
where $Y_{\rm L}(\eta)$ and $Y_{\rm T} (r,b)$ are longitudinal (Gaussian) and transverse (linear) flow rapidity 
profiles respectively, $R(b,\phi)$ is the fireball transverse radius in the azimuthal direction 
$\phi$, and $\delta(b)$ is the momentum anisotropy parameter (the hydro-inspired pa\-ra\-met\-ri\-za\-ti\-on 
~\cite{Wiedemann:1997cr} is implemented in HYDJET++ for the momentum and spatial anisotropy of a 
thermal hadron emission source). Since $V_{\rm eff}$ is a functional of the field of collective velocities on 
the freeze-out hypersurface, in fact the hadron spectrum is constructed as the superposition of 
thermal distribution and collective flow. The simulation procedure includes generation of a 
hadron four-momentum in the liquid element rest frame in accordance with the equilibrium distribution
function, generation of a spatial position and local four-velocity of the liquid element in accordance 
with the phase space and the character of fluid motion, boost of the hadron four-momentum in the event 
center-of-mass frame, and finally two- and three-body decays of resonances with branching ratios taken from 
the SHARE particle decay table~\cite{Torrieri:2004zz}. At first the value $V_{\rm eff}$ is calculated for central 
collisions ($b=0$), and then for non-central collisions it is supposed to be proportional to the mean 
number of nucleons-participants at given $b$. The event-by-event simulation of hadron production assumes 
the Poisson multiplicity distribution around its mean value for each hadron species.  

The scenario with different chemical and thermal freeze-outs is implemented in HYDJET++. It means that 
particle number ratios are fixed at chemical freeze-out temperature $T^{\rm ch}$, while the effective 
thermal volume $V_{\rm eff}$ and hadron momentum spectra being computed at thermal freeze-out temperature 
$T^{\rm th} \le T^{\rm ch}$. Introducing the temperature of chemical freeze-out (lower than hadronization 
temperature $T_{\rm c}$) and the temperature of thermal freeze-out effectively trace the stages of inelastic 
(between $T_{\rm c}$ and $T^{\rm ch}$) and elastic (between  $T^{\rm ch}$ and $T^{\rm th}$) hadronic 
rescatterings. Thus lower value of $T^{\rm th}$ with respect to $T^{\rm ch}$ would indicate on separate 
chemical and thermal freeze-out for given hadrons species. Such simplified but fast freeze-out approach is 
different from the simulation of full hadron cascade evolution requiring huge computing efforts.  

\subsection {Hard component} 

The approach for the hard component is based on the PYQUEN jet quenching 
model~\cite{Lokhtin:2005px} modifying the nucleon-nucleon collisions generated with PYTHIA$\_$6.4 
event generator~\cite{Sjostrand:2006za}. The basic kinetic equation for 
partonic energy loss $\Delta E$ as a function of initial energy $E$ 
and path length $L$ has the form: 
\begin{eqnarray} 
\label{elos_kin}
\Delta E (L,E) = \int\limits_0^Ldl \frac{dE(l,E)}{dl} \exp{\left( -l/\lambda(l)\right) }~,
\end{eqnarray} 
where $l$ is the current transverse coordinate of a parton, $dE/dl$ is the energy loss 
per unit length, $\lambda = 1/(\sigma \rho)$ is the in-medium mean free path, 
$\rho \propto T^3$ is the medium density at the temperature $T$, $\sigma$ is the 
integral cross section for the parton interaction in the medium. 

The radiative energy loss of massless quark is computed within BDMPS 
model~\cite{Baier:1996kr,Baier:1999ds,Baier:2001qw} as
\begin{eqnarray} 
\label{radiat} 
& & \frac{dE}{dl}^{\rm rad} = \frac{2 \alpha_s \mu_{\rm D}^2C_{\rm R}}{\pi L}
\int\limits_{\mu_{\rm D}^2\lambda_{\rm g}}^E  
d \omega \left[ 1 - x + \frac{x^2}{2} \right] 
\>\ln{\left| \cos{(\omega_1\tau_1)} \right|}~, \nonumber \\ 
& & \omega_1 = \sqrt{i \left( 1 - x + \frac{C_{\rm R}}{3}x^2 \right)   
\bar{\kappa}\ln{\frac{16}{\bar{\kappa}}}}~,~~ 
\bar{\kappa} = \frac{\mu_{\rm D}^2\lambda_{\rm g}}{\omega(1-x)} ~, 
\end{eqnarray} 
where $\tau_1=L/(2\lambda_{\rm g})$, $x=\omega/E$ is the fraction of the quark  
energy carried away by the radiated gluon, $\alpha_{\rm s}$ is the QCD running 
coupling constant for $N_{\rm f}$ active quark flavors, $C_{\rm R} = 4/3$ is the quark 
color factor, and $\mu_{\rm D}$ is the Debye screening mass.  
The simple generalization of the formula (\ref{radiat}) for a heavy quark of mass 
$m_{\rm q}$ is based on the ``dead-cone'' approximation~\cite{Dokshitzer:2001zm}:  
\begin{equation}
\label{radmass} 
\frac{dE}{dl}^{\rm rad}| _{m_{\rm q} \ne 0} =  \frac{1}{(1+(\beta \omega )^{3/2})^2}
\frac{dE}{dl}^{\rm rad}| _{m_{\rm q}=0}~, ~~~
\beta =\left( \frac{\lambda}{\mu_{\rm D}^2}\right) 
^{1/3} \left( \frac{m_{\rm q}}{E}\right) ^{4/3}~. 
\end{equation}

The collisional energy loss due to elastic scatterings is calculated in the 
high-momentum transfer limit~\cite{Bjorken:1982tu,Braaten:1991jj,Lokhtin:2000wm}: 
\begin{eqnarray}
\label{col} 
\frac{dE}{dl}^{\rm col} = \frac{1}{4T \lambda \sigma} 
\int\limits_{\displaystyle \mu^2_{\rm D}}^
{\displaystyle t_{\rm max}}dt\frac{d\sigma }{dt}t \, ,
\end{eqnarray} 
where $t$ is the momentum transfer square, and the dominant contribution to the differential scattering cross section is 
\begin{eqnarray} 
\label{sigt} 
\frac{d\sigma }{dt} \cong C \frac{2\pi\alpha_{\rm s}^2(t)}{t^2} 
\frac{E^2}{E^2-m_{\rm q}^2} \, 
\end{eqnarray} 
for the scattering of a hard quark with energy $E$ and mass $m_{\rm q}$ off the ``thermal'' 
parton with energy $m_0 \sim 3T \ll E$, $C = 1$ and $4/9$ 
for $qg$ and $qq$ scatterings respectively. The integrated cross section $\sigma$ 
is regularized by the Debye screening mass squared $\mu_D^2 (T) \simeq 4\pi 
\alpha _s T^2(1+N_f/6)$. The maximum momentum transfer square is 
$t_{\rm max}=[ s-(m_{\rm q}+m_0)^2] [ s-(m_{\rm q}-m_0)^2 ] / s$ where $s=2m_0E+m_0^2+m_{\rm q}^2$. 

Note that a number of more recent and sophisticated developments in partonic energy 
loss calculations (for massless partons as well as for heavy quarks) are available 
in the literature (see, e.g.,~\cite{Qin:2015srf} for the overview). For example, our simplification is that the collisional energy loss 
due to elastic scatterings with low momentum transfer (resulting mainly from the 
interactions with quark-gluon plasma collective modes or color background fields) is 
not taken into account. In principle, in the majority of estimations, latter process 
does not contribute much to the total collisional loss in comparison with the 
high-momentum scattering, and in numerical computations it can be effectively 
``absorbed'' by means of redefinition of minimum momentum transfer used to regularize 
the integral elastic cross section. Anyway, the treatment of radiative and collisional 
energy loss within HYDJET++ is used in current study mainly just to illustrate the influence of 
non-thermal component on charm production features. 

The medium where partonic rescattering occurs is treated as a boost-invariant 
longitudinally expanding perfect quark-gluon fluid, and the partons as being produced on a 
hyper-surface of equal proper times $\tau$~~\cite{bjork86}. The strength of partonic energy 
loss in PYQUEN is determined mainly by the initial maximal temperature $T_0^{\rm max}$ of the 
hot fireball in central PbPb collisions, which is achieved in the center of nuclear 
overlapping area at mid-rapidity. The transverse energy density in 
each point inside the nuclear overlapping zone is supposed to be proportional to the 
impact-parameter dependent product of two nuclear thickness functions $T_A$ in this point. 

Monte-Carlo simulation procedure in PYQUEN includes generation of 
the initial parton spectra with PYTHIA and production vertices at given impact 
parameter, rescattering-by-rescattering simulation of the parton path in a 
dense zone, radiative and collisional energy loss per rescattering, final 
hadronization with the Lund string model for hard partons and in-medium emitted 
gluons. The PYQUEN multi-jet state is generated according to the binomial distribution, and
then included in the hard component of HYDJET++ event. The mean number of jets (including heavy 
quark pairs) produced in AA events at a given impact parameter $b$ is computed as
\begin{eqnarray}
\label{numjets}
\overline{N_{\rm AA}^{{\rm jet}}} (b,\sqrt{s},p_{\rm T}^{\rm min}) =      
\int\limits_{p_{\rm T}^{\rm min}} dp_{\rm T}^2 \int dy  
\frac{d\sigma_{\rm NN}^{\rm hard}(p_{\rm T},~\sqrt{s})}{dp_{\rm T}^2dy}
\nonumber \\
\int\limits_0^{2\pi}d \psi \int\limits_0^{\infty }rdr 
T_A(r_1)T_A(r_2)S(r_1,r_2,p_T,y)~, 
\end{eqnarray} 
where $d\sigma_{\rm NN}^{\rm hard}(p_T, \sqrt{s})/dp_T^2dy$, 
calculated with PYTHIA, is the 
differential cross section of the hard process in NN collisions with the minimum 
transverse momentum transfer $p_T^{\rm min}$, and  $r_{1,2}$ are the transverse distances between the centres of colliding nuclei and the jet production vertex. The partons produced in hard processes 
with the momentum transfer lower than $p_{\rm T}^{\rm min}$ are considered as ``thermalized''. 
So, their hadronization products (including $D$ and $J/\psi$  mesons) are presented ``automatically'' in the soft component.
The contribution of hard component into the total multiplicity is controlled by the 
parameter $p_{\rm T}^{\rm min}$. The dominant contribution into charmed meson and charmonim yields comes from the soft component, the hard component being important at high $p_{\rm T}$. The prompt and non-prompt $J/\psi$ fractions for the hard component are taken from PYTHIA with subsequent modification of non-prompt $J/\psi$ meson spectra due to medium-induced b-quark energy loss simulated via PYQUEN. 
The factor $S \le 1$  in~(\ref{numjets}) takes into account the effect of nuclear shadowing on parton 
distribution functions (PDF). It is computed using the impact parameter dependent 
pa\-ra\-met\-ri\-za\-ti\-on~\cite{Tywoniuk:2007xy} obtained in the framework of Glauber-Gribov theory. Note that nuclear shadowing corrections are introduced only for hard component, while the soft component modeling does not use PDF explicitly. 

\subsection{Charmed mesons and charmonia at RHIC}

The input parameters of HYDJET++ for soft and hard components have been tuned from fitting
to heavy ion data on various observables for inclusive hadrons at RHIC~\cite{Lokhtin:2008xi} 
and LHC~\cite{Lokhtin:2012re}. 

It was shown in~\cite{Lokhtin:2010ze} that using the same values for the $J/\psi$ thermal and chemical freeze-out temperatures (with reduced radial and longitudinal collective velocities) allows HYDJET++ to properly reproduce 
$p_{\rm T}$- and $y$-spectra measured by PHENIX 
Collaboration~\cite{Adare:2006ns} in central AuAu collisions at RHIC energy $\sqrt{s_{\rm NN}}=200$ GeV. Note that the early thermal freeze-out of $J/\psi$-mesons was already suggested some years ago to describe SPS PbPb data at 
beam energy 158 GeV/nucleon~\cite{Bugaev:2001sj}. One may argue that this is due to the
higher mass and lower interaction cross section of the heavy  mesons.
We also have checked that $p_{\rm T}$-spectrum of $D$-mesons measured by STAR Collaboration~\cite{Adamczyk:2014uip} in 
central AuAu collisions at $\sqrt{s_{\rm NN}}=200$ GeV is reproduced by HYDJET++ simulation with 
the same freeze-out parameters as for $J/\psi$-mesons, but not for inclusive hadrons. It means  
that $D$-mesons like $J/\psi$-mesons are not in a kinetic equilibrium with the created medium at RHIC. 
Then let us get a look at the LHC situation. 

\section{$J/\psi$-meson production in lead-lead collisions at $\sqrt{s_{\rm NN}}=2.76$ TeV}

As it was already mentioned in previous section, the input parameters of HYDJET++  have been tuned from fitting 
to PbPb data at $\sqrt{s_{\rm NN}}=2.76$ TeV for inclusive hadrons~\cite{Lokhtin:2012re}. The most important 
parameters for our current consideration are the chemical and thermal freeze-out temperatures, 
$T_{\rm ch}=165$ MeV and $T_{\rm th}=105$ MeV, maximal longitudinal and transverse flow rapidities, 
$Y_{\rm L}^{\rm max}=4.5$ and $Y_{\rm T}^{\rm max}=1.265$, minimal transverse momentum
transfer of initial hard scatterings $p_{\rm T}^{\rm min}=8.2$ GeV/$c$, and initial maximal temperature 
of quark-gluon fluid $T_0^{\rm max}=1$ GeV. PYTHIA$\_$6.4 tune Pro-Q20 has been used to 
simulate an initial partonic state of the hard component. This tune reproduces the LHC data on inclusive 
hadron momentum spectra in pp collisions with the 10--15\% accuracy in the full measured
$p_{\rm T}$-range~\cite{Chatrchyan:2011av}.

\begin{figure}[h]
\center  \includegraphics[width=30pc]{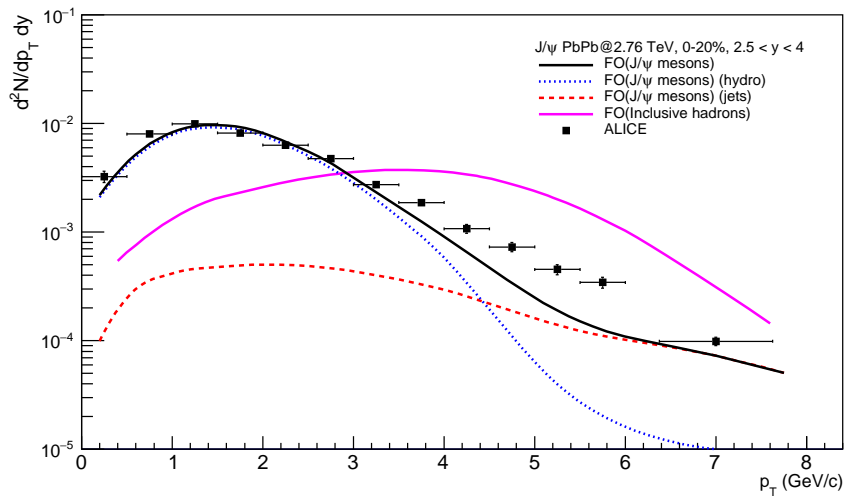}
		\caption{\label{jpsi_temp}
			Transverse momentum spectrum of inclusive $J/\psi$-mesons for rapidity $2.5<y<4$ in  
			20\% of most central PbPb collisions at $\sqrt{s_{\rm NN}}=2.76$ TeV. The points denote ALICE 
			data~\cite{Adam:2015isa}, histograms represent simulated HYDJET++ events (magenta solid -- freeze-out
			parameters as for inclusive hadrons, black solid -- early thermal freeze-out, blue dotted and red dashed -- soft and hard components respectively for the latter case).}
\end{figure}

\begin{figure}[h]
	\center	\includegraphics[width=30pc]{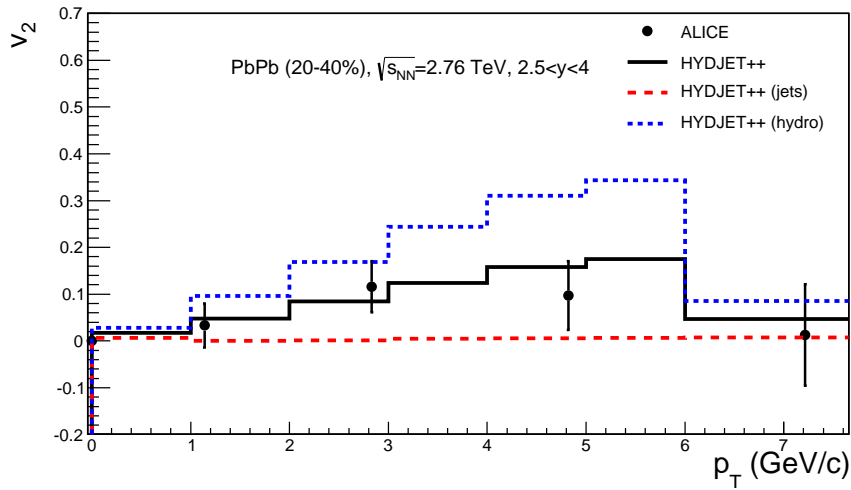}
		\caption{\label{jpsi_v2} Elliptic flow coefficient $v_2(p_{\rm T})$ of inclusive $J/\psi$-mesons for rapidity
			$2.5<y<4$ in the 20--40\% centrality class of PbPb collisions at $\sqrt s_{\rm NN}=2.76$ TeV. 
			The points denote ALICE data~\cite{ALICE:2013xna}, histograms represent simulated HYDJET++ events  
			(blue dotted -- soft component, red dashed -- hard component, black solid - both components).} 
\end{figure}

Figure~\ref{jpsi_temp} shows 
the comparison of HYDJET++ simulations with the ALICE data~\cite{Adam:2015isa} for $p_{\rm T}$-spectrum 
of inclusive $J/\psi$-mesons in 20\% of most central PbPb collisions at $\sqrt{s_{\rm NN}}=2.76$ TeV with two sets 
of input parameters: {\it 1)} as for inclusive hadrons (listed above), and {\it 2)} for early thermal 
freeze-out ($T_{\rm ch}=T_{\rm th}=165$ MeV, $Y_{\rm L}^{\rm max}=2.3$, $Y_{\rm T}^{\rm max}=0.6$, 
$p_{\rm T}^{\rm min}=3.0$ GeV/$c$). The fugacity value $\gamma_{\rm c} = 11.5$ was fixed from absolute 
$J/\psi$ yields. One can see that the situation is similar to RHIC: simulated spectra match the data 
(up to $p_{\rm T} \sim 3$ GeV/$c$) only assuming early thermal freeze-out, which happens  
presumably at the phase of chemical freeze-out (with reduced collective velocities, and enhanced contribution of 
non-thermal component). Note that we did not tune PYTHIA specially for charmonium production. This is the important but rather specific task which is out of the scope of current paper. We have checked that the PYTHIA version used for our simulations indeed underestimates
the charmonium yield measured by ALICE in pp collisions at $\sqrt{s}=2.76$ TeV~\cite{Abelev:2012kr}. This discrepancy is responsible for the disagreement between HYDJET++ and the PbPb data at high $p_{\rm T}$, where the contribution from hard component becomes significant. 

In addition, we found that the $p_{\rm T}$-dependence of the elliptic flow coefficient $v_2$ measured by 
ALICE for inclusive $J/\psi$'s~\cite{ALICE:2013xna} is reproduced by HYDJET++ simulation in the case of  
early freeze-out (figure~\ref{jpsi_v2}). An important role of non-thermal component at high $p_{\rm T}$ is clearly seen.

Thus we conclude that the significant part of $J/\psi$-mesons (up to $p_{\rm T} \sim 3$ GeV) produced in PbPb collisions 
at the LHC is thermalized, but without being in kinetic equilibrium with the medium (similar to the RHIC case). 

\section{$D$-meson production in lead-lead collisions at $\sqrt{s_{\rm NN}}=2.76$ TeV}

\begin{figure}[h]
	\center	\includegraphics[width=30pc]{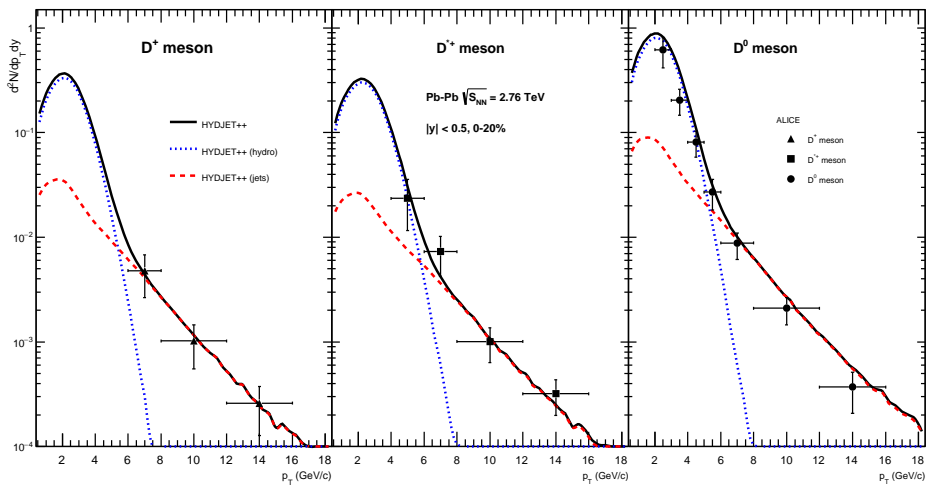}
		\caption{\label{d_pt}
			Transverse momentum spectra of $D^{\pm}$ (left panel), $D^{*\pm}$ (middle panel) and $D^0$ (right panel) for rapidity $\mid y \mid < 0.5$
			in 20\% of most central PbPb collisions at $\sqrt{s_{\rm NN}}=2.76$ TeV. The points denote ALICE 
			data~\cite{ALICE:2012ab}, histograms represent simulated HYDJET++ events (blue dotted -- soft component, red dashed -- hard component, black solid - both components).}
\end{figure}

At first, we simulated $D$-meson production with the same freeze-out parameters as for inclusive hadrons.
The fugacity value $\gamma_c=11.5$ was fixed from $J/\psi$ yield. 
Figure~\ref{d_pt} shows the comparison of HYDJET++ simulations with the ALICE data~\cite{ALICE:2012ab} for 
$p_{\rm T}$-spectra of $D^{\pm}$, $D^{*\pm}$ and $D^0$ mesons in 20\% of most central PbPb collisions 
at $\sqrt{s_{\rm NN}}=2.76$ TeV. The simulated results are close to the data within the experimental  
uncertainties. Thus in contrast to RHIC, thermal freeze-out of $D$-mesons at the LHC happens
simultaneously with thermal freeze-out of light hadrons. 

\begin{figure}[h]
\center	\includegraphics[width=30pc]{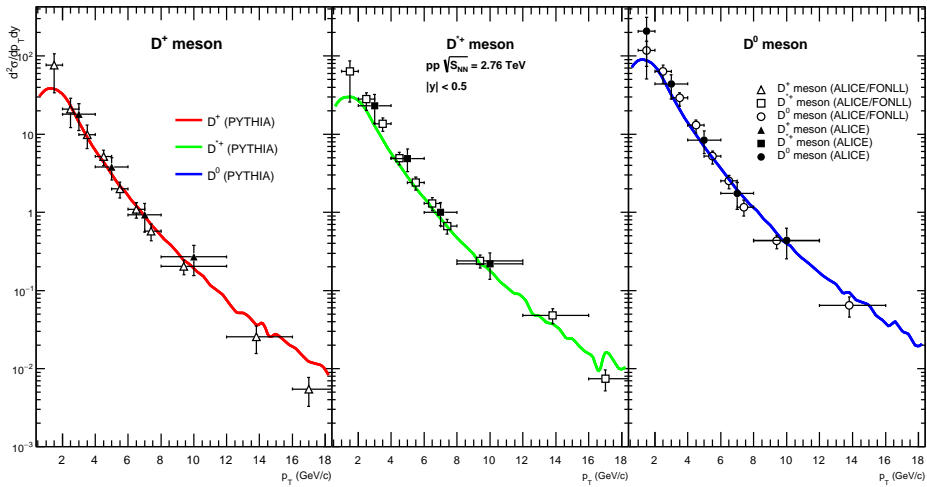}
		\caption{\label{d_pp}
			Transverse momentum spectra of $D^{\pm}$ (left panel), $D^{*\pm}$ (middle panel) and $D^0$ (right panel) for rapidity $\mid y \mid < 0.5$ in pp collisions at $\sqrt{s_{\rm NN}}=2.76$ TeV. The closed and open points denote ALICE 
			data~\cite{Abelev:2012vra} and FONLL-based extrapolation of pp data from $\sqrt{s}=7$ TeV respectively, histograms 
represent simulated PYTHIA$\_$6.4 events.}
\end{figure}

\begin{figure}[h]
	\center	\includegraphics[width=30pc]{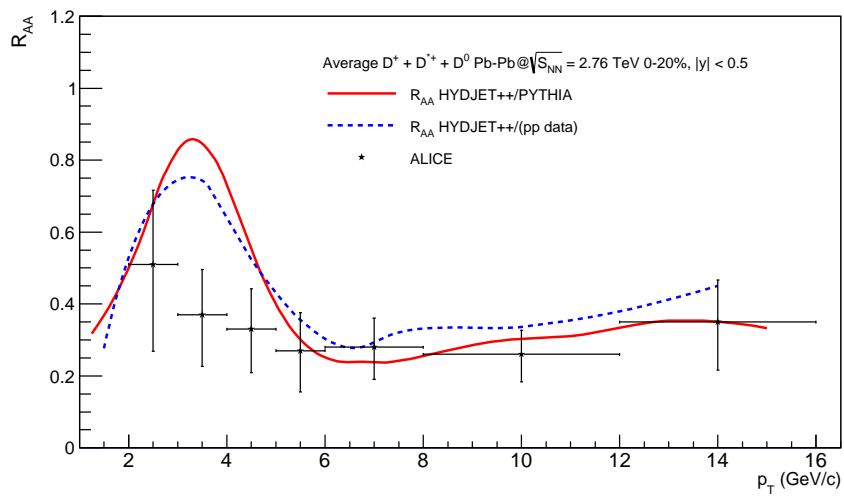}
		\caption{\label{d_raa}
                 Average of the three $D$-meson species nuclear modification factor  $R_{\rm AA}(p_{\rm T})$ for rapidity $\mid y \mid < 0.5$
			in 20\% of most central PbPb collisions at $\sqrt{s_{\rm NN}}=2.76$ TeV. The points denote ALICE 
			data~\cite{ALICE:2012ab}, histograms represent simulated HYDJET++ events for two pp references 
(blue dashed -- extrapolated pp data, red solid -- PYTHIA).}
\end{figure}

\begin{figure}[h]
	\center	\includegraphics[width=30pc]{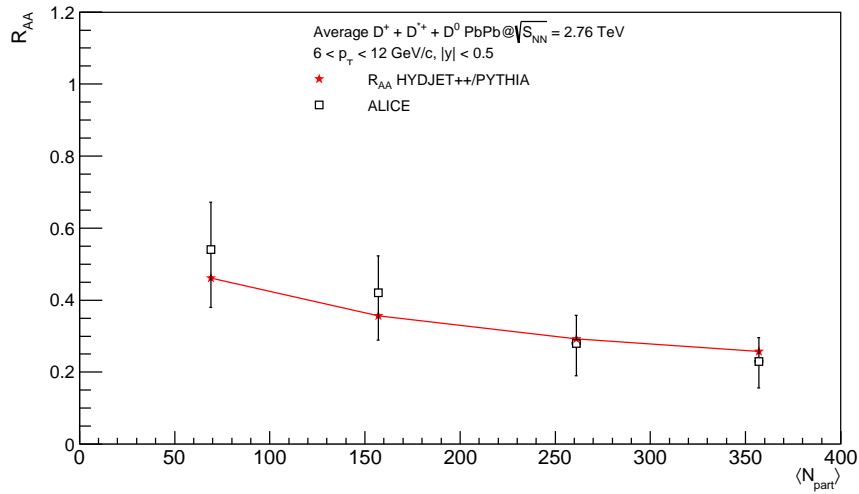}
		\caption{\label{d_raa_npart}
                 Centrality dependence of average of the three $D$-meson species nuclear modification factor $R_{\rm AA}$ 
                for rapidity $\mid y \mid < 0.5$ and $6 < p_{\rm T} < 12$ GeV/$c$ in PbPb collisions 
                at $\sqrt{s_{\rm NN}}=2.76$ TeV. The open squares denote ALICE data~\cite{ALICE:2012ab}, 
                asterisks represent simulated HYDJET++ events. The line is drawn to guide the eye.}
\end{figure}

\begin{figure}[h]
           \center   \includegraphics[width=30pc]{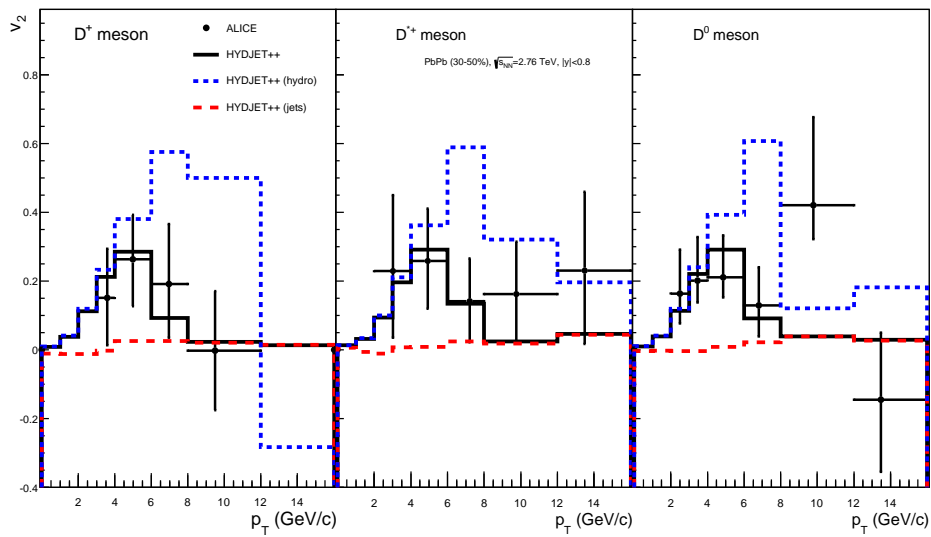}
		\caption{\label{d3_v2} 
			Elliptic flow coefficient $v_2(p_{\rm T})$ of $D^{\pm}$ (left panel), $D^{*\pm}$ (middle panel) 
                        and $D^0$ (right panel) mesons at rapidity
			$\mid y \mid <0.8$ in the 30--50\% centrality class of PbPb collisions at 
			$\sqrt s_{\rm NN}=2.76$ TeV. The points denote ALICE data~\cite{Abelev:2013lca}, 
                        histograms represent  simulated HYDJET++ events (blue dotted -- soft component, 
                        red dashed -- hard component, black solid - both components).}
\end{figure}

\begin{figure}[h]
    \center    \includegraphics[width=30pc]{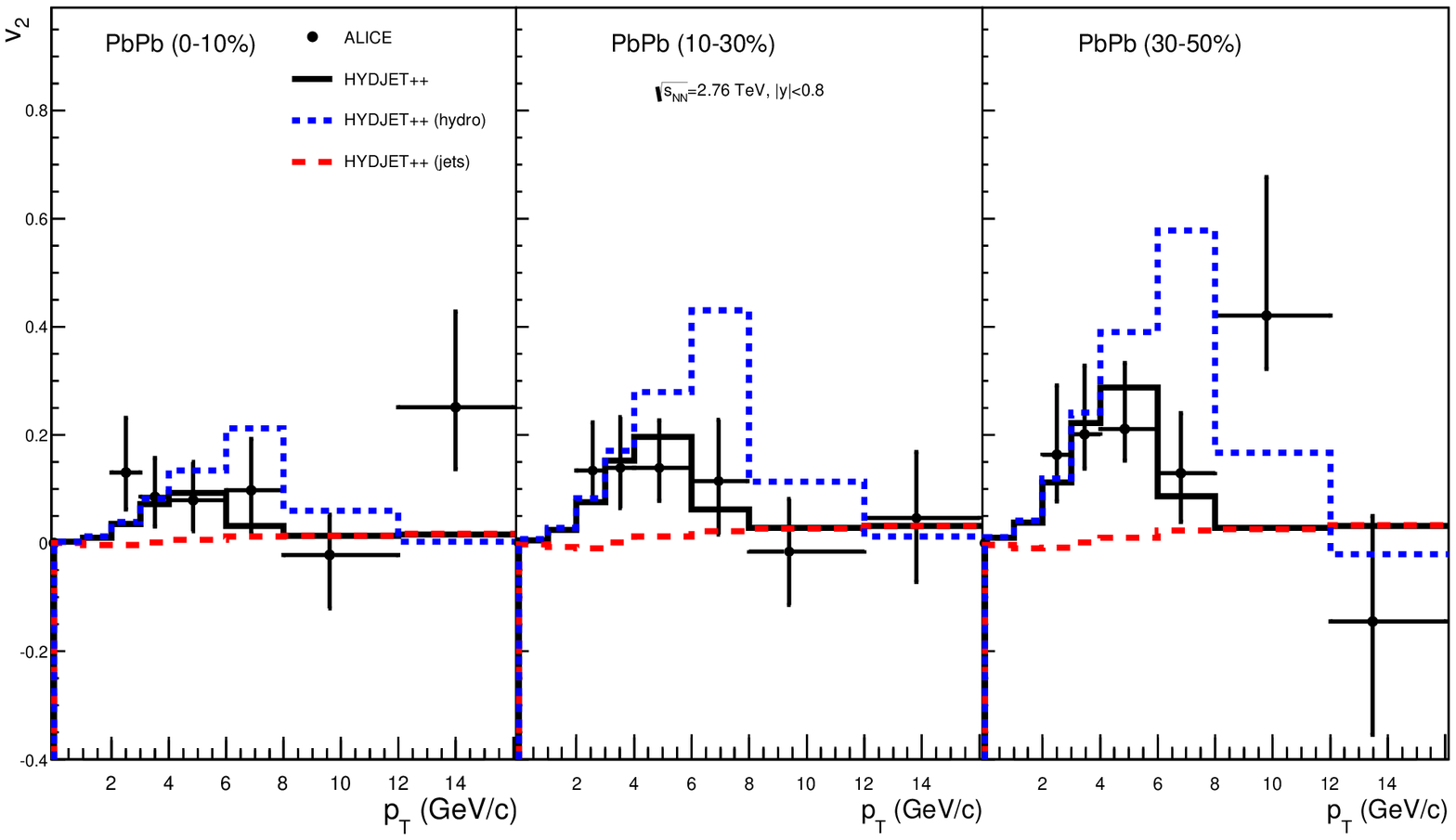}
		\caption{\label{d_v2} 
			Elliptic flow coefficient $v_2(p_{\rm T})$ of $D^0$-mesons at rapidity
			$\mid y \mid <0.8$ in the 0--10\% (left panel), 10--30\% (middle panel) and 30--50\% 
                        (right panel) centrality classes of PbPb collisions at $\sqrt s_{\rm NN}=2.76$ TeV. 
                        The points denote ALICE data~\cite{Abelev:2014ipa}, histograms represent  simulated 
                        HYDJET++ events (blue dotted -- soft component, red dashed -- hard component, 
                        black solid - both components).}
\end{figure}

The high-$p_{\rm T}$ particle production in heavy ion collisions is characterized by the
nuclear modification factor $R_{\rm AA}$, which is defined as a ratio of particle yields in AA and pp collisions normalized on 
the mean number of binary nucleon-nucleon sub-collisions $\left<N_{\rm coll}\right>$ for given event centrality class 
(calculated within 
HYDJET++):
\begin{equation}
\label{raa}
R_{\rm AA}(p_{\rm T})=\frac{d^2N^{\rm AA}/dy dp_{\rm T}}
{\left<N_{\rm coll}\right>d^2N^{\rm pp}/dy dp_{\rm T}}~.
\end{equation}
In the absence of nuclear effects (in initial or final states) at high $p_{\rm T}$, it should be $R_{\rm AA}=1$.

To estimate the uncertainties related to choice of pp reference needed for the construction of $R_{\rm AA}$~(\ref{raa}) for 
$D$-mesons, we compare the predictions from PYTHIA$\_$6.4 (tune Pro-Q20) and 
from FONLL-based extrapolation of pp data from $\sqrt{s}=7$ TeV. The latter procedure was utilized by ALICE Collaboration 
in~\cite{ALICE:2012ab}. It includes determination of the reference pp cross sections for each D-meson species applying 
$\sqrt{s}$-scaling~\cite{Averbeck:2011ga} based on FONLL calculations~\cite{Cacciari:1998it,Cacciari:2001td,Cacciari:2012ny}
to the cross sections measured at $\sqrt{s}=7$ TeV~\cite{ALICE:2011aa}. 
We have found that D-meson $p_{\rm T}$-spectra at $\sqrt{s}=2.76$ TeV for both pp references are similar, and close to the 
experimental data~\cite{Abelev:2012vra} (figure~\ref{d_pp}). 

The $p_{\rm T}$-dependence of average of the three $D$-meson species nuclear modification factor $R_{\rm AA}$ is presented for both pp 
references on figure~\ref{d_raa}. The simulated results are close to the data up to highest $p_{\rm T}=16$ GeV. 
The measured centrality dependence of nuclear modification factor for high-$p_{\rm T}$ $D$-mesons~\cite{ALICE:2012ab} is described well 
by HYDJET++ simulations. This is demonstrated in figure~\ref{d_raa_npart}, where the events are divided in four centrality 
classes (0--10\%, 10--20\%, 20--40\% and 40--60\%), and characterized by the average number of participating 
nucleons $\langle N_{\rm part} \rangle $. In this case, the PYTHIA pp reference was used for the denominator of the simulated $R_{\rm AA}$.
 
Finally, HYDJET++ is able to reproduce the ALICE data ~\cite{Abelev:2013lca,Abelev:2014ipa} on $p_{\rm T}$-dependence of the elliptic flow 
coefficient $v_2$ (figures~\ref{d3_v2},\ref{d_v2}). Thus the simultaneous description of $p_{\rm T}$-spectrum and elliptic flow coefficients of charmed mesons is achieved within the model.

We conclude that the significant part of $D$-mesons (up to $p_{\rm T} \sim 4$ GeV/$c$) produced in PbPb collisions 
at the LHC seems to be in a kinetic equilibrium with the medium. This is quite different from the RHIC situation. The possible reason 
for this may be that $D$-meson interaction cross section at LHC energy becomes comparable with the interaction cross section of light hadrons, 
but $J/\psi$-meson interaction cross section remains much smaller. The momentum and centrality dependencies of nuclear 
modification factor for $D$-mesons at high-$p_{\rm T}$ are reproduced by HYDJET++ modeling.  

\section{Summary}

The phenomenological analysis of charmed meson and charmonium production in 
lead-lead collisions at the center-of-mass energy 2.76 TeV per nucleon pair has been done 
within the two-component HYDJET++ model including thermal and non-thermal production mechanisms. 
Momentum spectra and elliptic flow of $D$ and $J/\psi$ mesons are simultaneously reproduced by 
the model assuming that thermal freeze-out of $D$-mesons happens simultaneously with thermal 
freeze-out of light hadrons, while thermal freeze-out of $J/\psi$-mesons happens appreciably before,
presumably at the phase of chemical freeze-out (with reduced radial and longitudinal collective velocities). 
Thus the significant part of $D$-mesons (up to transverse momenta $p_{\rm T} \sim 4$ GeV/c), unlike 
$J/\psi$ mesons, seems to be in a kinetic equilibrium with the created in PbPb collisions hot 
hadronic matter. It may indicate that D-meson interaction cross section at the LHC becomes comparable 
with the interaction cross section of light hadrons, but $J/\psi$-meson interaction cross section remains 
significantly smaller.

Non-thermal charm production mechanism is important at high transverse momenta. A good agreement of the simulated results with the data for $D$-meson nuclear modification factors  
testifies in favor of successful treatment of hard charm component within the framework of the HYDJET++ model. 

\ack
Discussions with J.~Bielcik, L.V.~Bravina, A.I.~Demianov, V.L.~Korotkikh, L.V.~Malinina, A.M.~Snigirev, E.E.~Zabrodin 
and S.V.~Petrushanko are gratefully acknowledged. We thank our colleagues from ALICE and CMS 
collaborations for fruitful cooperation. This work was supported by the Russian Science Foundation  
under Grant No. 14-12-00110 in a part of computer simulation of $p_{\rm T}$-spectrum and elliptic flow of $D$ and $J/\psi$ 
mesons in lead-lead collisions.

\end{document}